# Microscopic magnetization distribution of Bloch lines in a uniaxial magnet

K. Kurushima[1], K. Tanaka[2], H. Nakajima[3], M. Mochizuki[2], and S. Mori[3]

[1]*Toray Research Center, Ohtsu, Shiga 520-8567, Japan*

[2]*Department of Applied Physics, Waseda University, Okubo, Shinjuku-ku, Tokyo 169-8555, Japan*

[3]*Department of Materials Science, Osaka Prefecture University, Sakai, Osaka 599-8531, Japan*

Bloch lines are formed to reduce the magnetostatic energy generated by the Bloch walls in uniaxial magnets. Recently, it is reported that Bloch lines play important roles in the emergence and helicity reversal of magnetic bubbles in Sc-substitute *M*-type hexaferrites ($BaFe_{12-x-0.05}Sc_xMg_{0.05}O_{19}$). Although Bloch lines have been discussed on the basis of micromagnetic simulations, the detailed structure was not observed directly. In this study, we investigated the microscopic structures of Bloch lines in $BaFe_{10.35}Sc_{1.6}Mg_{0.05}O_{19}$ uniaxial magnets. Differential-phase contrast scanning transmission microscopy (DPC-STEM) directly revealed that the edges of the Bloch walls were misaligned in the Bloch lines of $BaFe_{10.35}Sc_{1.6}Mg_{0.05}O_{19}$. From the micromagnetic simulations based on the Monte-Carlo technique, we showed that the misaligned Bloch walls were caused by the dipole-dipole interactions in the hexaferrite. Our results will help to understand the microstructures of Bloch lines at nanometer scale.

## I. INTRODUCTION

Uniaxial magnets are magnetic materials that show net magnetization along one particular crystallographic direction (easy axis). These magnets are used as permanent magnets and are essential in electronic appliances. Since their discovery, *M*-type hexaferrites ($BaFe_{12}O_{19}$) have been one of the most widely used permanent magnets.[1] Recently, many novel physical properties have been reported in nonmagnetic element-substituted *M*-type hexaferrites. Sc-substituted *M*-type hexaferrites (BaFe$_{12-x-0.05}$Sc$_x$Mg$_{0.05}$O$_{19}$) exhibit magnetoelectric effects owing to their spiral magnetic structures (longitudinal conical magnetic structures).[2] The conical magnetic structure depends on the amount of substituted Sc; the conical transition temperature exceeds room temperature at $x > 1.8$, which allows *M*-type hexaferrites to be applied in magnetic devices.[3,4]

Another important property of Sc-substituted *M*-type hexaferrite is the peculiar behavior of magnetic bubbles.[5–7] Magnetic bubbles are vortex domain structures formed in uniaxial magnets as a result of the balance between magnetostatic, magnetic domain-wall, and Zeeman energies. Magnetic bubbles have recently attracted considerable interest[8–14] because of magnetic skyrmions,[8,15–21] which are nanometric vortex spin structures. In Sc-substituted *M*-type hexaferrite, the helicity (clockwise or counterclockwise) of magnetic bubbles is frequently reversed around the transition temperature.[6] Helicity is considered to be exchanged through Bloch lines, which are reverse-chirality points at Bloch walls (magnetic domain walls whose magnetizations rotate perpendicular to the walls). Lorentz microscopy observations further revealed that magnetic bubbles form by pinching off Bloch lines in hexaferrite because Bloch lines are energetically unfavorable.[7] These studies highlight the importance of Bloch lines in Sc-substituted hexaferrite and emphasize the need to study the microscopic structures of Bloch lines.

Bloch walls are divided into segments at Bloch lines, and the magnetization rotates within the Bloch walls. In hexaferrites with uniaxial anisotropy, the magnetizations at two Bloch walls point in the same direction (head-to-head or tail-to-tail structure). However, such head-to-head and tail-to-tail Bloch lines themselves are unstable, although Bloch lines are formed to reduce the stray field from Bloch walls.[22] Although several experiments have revealed the positions of Bloch lines,[23–26] their detailed microstructures remain elusive because available observation methods lack the resolution needed for direct observation. Thus, the microscopic structures of Bloch lines have been analyzed theoretically.[22] However, recent advancements in differential-phase contrast scanning transmission electron microscopy (DPC STEM)[18,27–29] have made it possible to directly observe magnetic domains including Bloch lines at high resolution.

In this paper, we report the microscopic structures of Bloch lines in the Sc-substituted $M$-type hexaferrite $BaFe_{12-x-0.05}Sc_xMg_{0.05}O_{19}$, $x = 1.6$ (BFSMO). High-resolution DPC STEM images directly show the substructures of the Bloch lines. The Bloch lines are revealed to comprise two misaligned Bloch walls. We reveal that the misaligned Bloch walls are caused by the dipole-dipole interactions. These structures are generated to reduce the magnetostatic energy induced by the Bloch lines. The edges of the Bloch walls are also shown to be misaligned in the Bloch line of a magnetic bubble.

## II. EXPERIMENTS

DPC STEM was performed using a transmission electron microscope (JEM-ARM200CF, JEOL Co. Ltd.) equipped with a segmented detector with eight segments.[30] The experimental setup is shown in Fig. 1(a). When electrons pass through a specimen possessing a magnetic field $B$, electrons are deflected by the Lorentz force, resulting in unequal intensities among the segments. The intensity is proportional to the magnetic field in the specimen, which is also proportional to the magnetization. Thus, the direction and magnitude of the magnetization can be directly deduced and mapped in real space based on the intensities.[31] The objective lens was turned off for magnetic domain observation and used to apply magnetic fields perpendicular to the thin specimen. DPC STEM images were taken under the in-focus condition. The probe convergence angle and detection angle were set to ~0.1 and ~50 mrad, respectively. In this setup, the outer detectors (5–8) had intensities from magnetically deflected electrons, whereas the inner detectors (1–4) had no intensity. Thus, the intensities of the inner detectors were used to subtract non-magnetic signals such as bend contours.[28] The resolution of DPC STEM in Lorentz mode was ~5 nm in this study. A single crystal of BFSMO was grown using a floating zone method. Thin specimens for DPC STEM observation were prepared using an $Ar^+$ ion-milling method. The thickness of the specimen was ~ 90 nm. The easy axis of BFSMO at room temperature is the $c$ axis; thus, the incident beam was set to be parallel to the $c$ axis of BFSMO.

Here, we briefly describe how we obtained the DPC STEM images from the experimentally obtained intensities (Figure 1b). The deflection angles along the $x$ and $y$ directions are proportional to the intensities $F_x = I_5 - I_7$ and $F_y = I_6 - I_8$, respectively, where $I_5$ through $I_8$ are the intensities in each segment. The magnetic field direction $\boldsymbol{B}$ is rotated by 90° with respect to the Lorentz force $\boldsymbol{F} = -e(\boldsymbol{v} \times \boldsymbol{B})$, where $e$ is the electric charge, and $\boldsymbol{v}$ is the velocity of electrons. $B_x = F_y$ and $B_y = -F_x$ were used to convert the direction of the Lorentz force to that of the magnetic field. The magnitude of the magnetic field (field strength image) was then calculated by $\sqrt{B_x^2 + B_y^2}$. In the color maps, the colors in each area were allocated by the angle ($\tan^{-1} B_y/B_x$) and the magnitude of the magnetic field. The upper panels in Fig. 2(b) show the intensity maps captured in each segment, while the lower panels show the magnetic field components ($B_x$ and $B_y$), field strength image, and color map of the Bloch walls. In addition, the rotation maps ($\frac{\partial B_y}{\partial x} - \frac{\partial B_x}{\partial y}$) were utilized to explicitly depict the helicity of the magnetic bubbles.

**III. RESULTS AND DISCUSSION**

Figure 2 shows the DPC STEM images of BFSMO at room temperature (field strength image and color maps). The field strength image indicates a stronger intensity in the domain walls compared to inside of the domains. In the color map, the direction and magnitude of magnetization are illustrated by color and contrast, respectively. Thus, the black contrast of the magnetic domains indicates that they had no in-plane magnetization component. Conversely, the magnetic domain walls were visualized by thin color lines, indicating in-plane magnetization components. These results are in accord with previous studies finding that magnetization points upward or downward in magnetic domains, and these domains are separated by Bloch walls.[6,7,32] It can be seen that Bloch walls were broken, as indicated by the white arrowheads (Bloch lines), and the intensities were low at the Bloch lines. To analyze these Bloch lines, we observed these areas at high magnification.

Figures 3(a) and 3(b) show high-magnification DPC STEM images around the Bloch lines. These images show that the Bloch wall width $w$ was ~50 nm [see Fig. 3(c)], whereas the Bloch line width (distance between points showing different colors) was ~120 nm. Therefore, the Bloch line width was larger than the Bloch wall width in

BFSMO, as expected in uniaxial magnets.[22] The Bloch wall width is comparable to a previous observation (wall width = 30 ~ 40 nm) of an *M*-type hexaferrite.[33] Considering that the magnetocrystalline anisotropy was reduced by the Sc substitution, the Bloch wall width is reasonable. In Fig. 3(a), the magnetizations pointed in the left and upper-right directions in the left and right Bloch walls, respectively. Thus, these walls formed a tail-to-tail magnetic domain structure. Noticeably, the edges of the Bloch walls were not aligned in a straight line; instead, they slightly deviated from each other. Furthermore, similar to in Fig. 3(a), a misaligned structure was observed in the head-to-head Bloch line [Fig. 3(b)]. These misaligned structures are considered to reduce the dipole-dipole interactions in the Bloch line.

Bloch walls are formed to decrease the magnetostatic energy (the stray field) of the bulk. Similarly, the Bloch line is formed to reduce the magnetostatic energy generated by the Bloch wall. However, Bloch lines themselves have high energy due to the divergence of the magnetic field if the magnetizations at the edges of the Bloch walls point in the opposite directions (head-to-head or tail-to-tail magnetic domain structures). Therefore, if the edges of the Bloch walls in the Bloch lines are misaligned, as observed in Fig. 3, the total energy is considered to be further reduced because of the reduction in magnetostatic energy induced by the Bloch lines. The misaligned structures decrease the stray field because the magnetizations with opposite directions are located in a close area similar to 180° domains.

In order to further clarify the physical origin of the observed misaligned Bloch lines, we performed numerical simulations. We first employ a continuum spin model for a thin-plate specimen of the *M*-type hexaferrite. The Hamiltonian is given by,

$$\mathcal{H} = A \int d\mathbf{r}(\nabla \mathbf{m})^2 + \frac{\mu_0 M_s^2}{4\pi} \int d\mathbf{r} \int d\mathbf{r}' \left\{ \frac{\mathbf{m}(\mathbf{r}) \cdot \mathbf{m}(\mathbf{r}')}{|\mathbf{r} - \mathbf{r}'|^3} + \frac{\mathbf{m}(\mathbf{r}) \cdot (\mathbf{r} - \mathbf{r}') \, \mathbf{m}(\mathbf{r}') \cdot (\mathbf{r} - \mathbf{r}')}{|\mathbf{r} - \mathbf{r}'|^5} \right\} - K_u \int d\mathbf{r} \, m_z^2$$

$$+ K_{\text{hex}} \int d\mathbf{r} \, (9 m_y^4 m_x^2 - 6 m_y^2 m_x^4 + m_y^6) \qquad (1)$$

where $\mathbf{m}(\mathbf{r})$ is the normalized classical magnetization vector. The first term describes the ferromagnetic exchange interactions with $A$ being the exchange stiffness, while the second term depicts the magnetic dipole-dipole interactions with $\mu_0$ and $M_s$ being the magnetic permeability for vaccum and the saturation magnetization, respectively. The third term denotes the perpendicular magnetic anisotropy with $z$ axis chose to be normal to the thin-plate plane. The fourth term describes the in-plane six-fold magnetic anisotropy due to the hexagonal crystalline symmetry of the *c*-plane of the *M*-type hexaferrites, which arranges hard (easy) magnetization axes shown in Fig. 4(a). The values of the parameters except for that of $K_{\text{hex}}$ are experimentally evaluated as $A = 1.3 \times 10^{-11}$ [J/m], $M_s = 2.9 \times 10^5$ [A/m], $K_u = 5.3 \times 10^4$ [J/m$^3$] on the basis of the magnetization measurement.[7] The exchange stiffness $A$ was calculated by using the equation $w = \pi\sqrt{A/K_u}$,[22,34] where $w = 50$ nm. We assume a plate-shaped sample of 960 nm × 720 nm × 96 nm.

Dividing the continuum space into cubic cells of $a^3$ with $a = 12$ nm, we obtained the following lattice spin model on the square lattice of $80 \times 60 \times 8$ sites as,

$$\mathcal{H} = -J \sum_{\langle i,j \rangle} \boldsymbol{m}_i \cdot \boldsymbol{m}_j$$

$$+ I_{\text{dip}} \sum_{(i,j)} \left\{ \frac{\boldsymbol{m}_i \cdot \boldsymbol{m}_j}{|\boldsymbol{i}-\boldsymbol{j}|^3} + \frac{\boldsymbol{m}_i \cdot (\boldsymbol{i}-\boldsymbol{j})\, \boldsymbol{m}_j \cdot (\boldsymbol{i}-\boldsymbol{j})}{|\boldsymbol{i}-\boldsymbol{j}|^5} \right\} - \kappa_u \sum_i m_{iz}^2$$

$$+ \kappa_{\text{hex}} \sum_i \left( 9 m_{iy}^4 m_{ix}^2 - 6 m_{iy}^2 m_{ix}^4 + m_{iy}^6 \right). \qquad (2)$$

Here the coupling coefficients $J = 2aA$, $I_{\text{dip}} = \mu_0 M_s^2 a^3/4\pi$, and $\kappa_u = K_u a^3$ are evaluated as $J = 1950$ meV, $\kappa_u = 572$ meV, and $I_{\text{dip}} = 88.4$ meV, respectively. When we set the energy unit as $J = 1$, we obtain $\kappa_u = 0.29$ and $I_{\text{dip}} = 0.045$. We assume $\kappa_{\text{hex}} = 0.03$.

In order to investigate stable magnetization structures of Bloch lines with this lattice spin model, we first prepared initial magnetization configurations shown in Figs. 4(b) and (c), in which the up-magnetization and the down-magnetization regions are partially separated by two lines. Each of these two lines runs along one of the easy magnetization axes determined by the in-plane magnetization anisotropy, and the magnetizations on each line are oriented along the line to form a head-to-head or tail-to-tail magnetization alignment. In addition, the magnetizations located in an area around which the two lines meet are randomly distributed initially. Starting with these initial configurations, we relaxed them by using the replica-exchange Monte-Carlo technique to explore stable structures of the Bloch lines at low temperature ($T/J = 0.01$). The open boundary conditions were imposed for all the simulations.

In the simulations, we obtained Bloch-line structures shown in Figs. 4(d) and (e), which clearly show the misalignments of two Bloch walls at their meeting point. We also examined the case without the magnetic dipole-dipole interaction by setting $I_{dip} = 0$. In this case, we obtained a normal Bloch line shown in Fig. 4(f) with no magnetization misalignment. If the dipole-dipole interactions do not exist, the magnetizations rotate gradually in a row, which agrees with other micromagnetic simulations.[25] These results indicate that the magnetic dipole-dipole interaction plays an essential role for the emergence of the misaligned Bloch lines through reducing the magnetostatic energy. In addition, by examining various cases with different parameters, we also found a crucial role of the in-plane six-fold magnetic anisotropy. The magnetizations in the normal Bloch line gradually change their orientations by 120° from one easy-axis direction to another easy-axis direction via the hard-axis directions. This magnetization alignment costs a large energy and thereby significantly destabilizes the normal Bloch-line structures, which gives a relative stability to the misaligned Bloch lines. This might be a reason why the specific misaligned Bloch line structures are observed in the present hexaferrite system.

Finally, we demonstrate that similar misalignment of Bloch wall edges is also realized in Bloch lines of type-II magnetic bubbles. By applying magnetic fields, the Bloch walls were moved, as shown in Figs. 5(a)–(d) [see the markers]. Furthermore, the Bloch walls were pinched off by the application of magnetic fields, and type-I and -II magnetic bubbles were formed. Here, type-I bubbles are defined as bubbles with continuously rotated Bloch walls, whereas type-II bubbles are defined as two parallel Bloch walls with two Bloch lines at their edges.[35] The rotation image [Fig. 5(d)] clearly shows the helicity of the magnetic bubbles. Figure 5(e) is a DPC STEM rotation image under a magnetic field of ~228 mT at high magnification around type- I and type-II magnetic bubbles. The type-II magnetic bubble (red rectangle) was observed at high magnification, as shown in Fig. 5(f). In the observed Bloch lines, Bloch walls were generated at different points, and the locations of the Bloch walls deviated at Bloch lines, similar to the case of the striped domain walls (Fig. 3). The existence of Bloch lines and similar misaligned Bloch walls can be seen in Figs. 4C and D of the previous work using the Fresnel method and a transport-of-intensity equation (TIE).[5] However, our DPC-STEM study clearly demonstrated the misaligned structure without the artifacts due to the Fresnel and TIE methods.[36] Therefore, we concluded that a similar misaligned structure of Bloch walls was realized in the Bloch lines of type-II magnetic bubbles.

## IV. CONCLUSIONS

We investigated the microstructures of Bloch lines at high resolution using DPC STEM. Our real-space observation clearly showed that the edges of Bloch walls were misaligned in Bloch lines. The micromagnetic simulations based on the Monte-Carlo technique demonstrated that the dipole-dipole interactions play an important role in the formation of the misaligned structure. This misalignment structure was also realized in the Bloch lines of type-II magnetic bubbles.

**ACKNOWLEDGEMENTS**

This work was partially supported by JSPS KAKENHI (Nos. 17H02924, 16H03833 and 15K13306) and a grant from the Murata Foundation. M. M. also thanks support from Waseda University Grant for Special Research Projects (Project No. 2017S-101) and JST PRESTO (Grant No. JPMJPR132A).

**Figures**

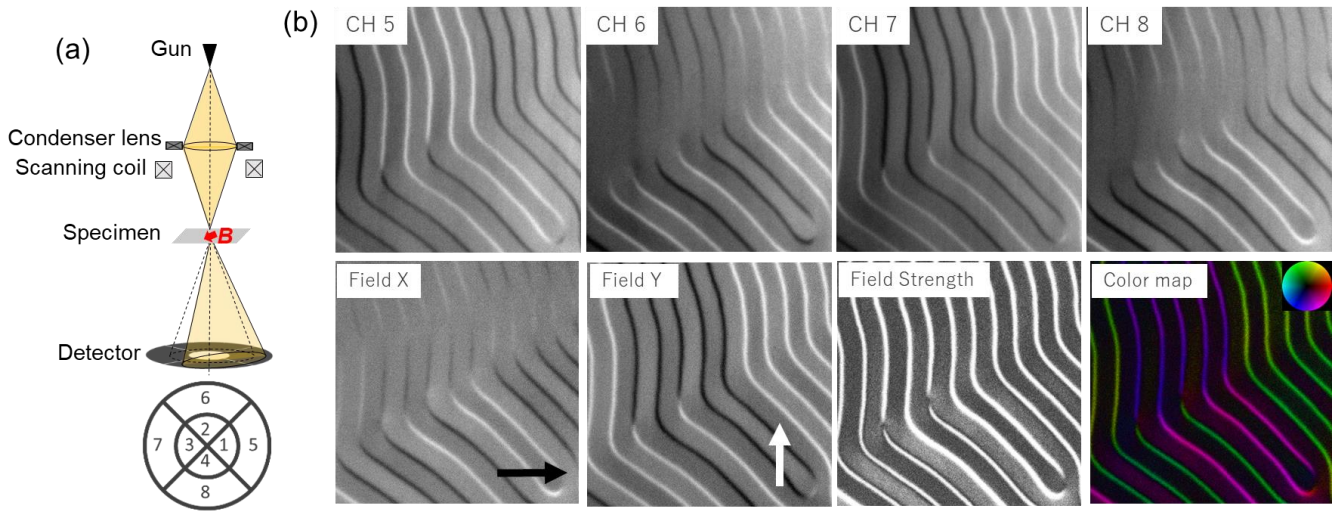

Fig. 1. (a) Experimental setup and principle of the DPC STEM imaging technique. The transmitted beam is deflected by the magnetic field *B* of the specimen. The detector segments are marked 1–8. (b) The experimental contrast of Bloch walls obtained using the outer detector 5–8 (upper) and the magnetic domain images calculated using the upper images (lower). Field X and Y represent the magnitudes of the magnetic fields along the *x* (black arrow) and *y* (white arrow) directions, respectively, which are proportional to the magnetization. The field strength image and color map are depicted based on the methods explained in the main text.

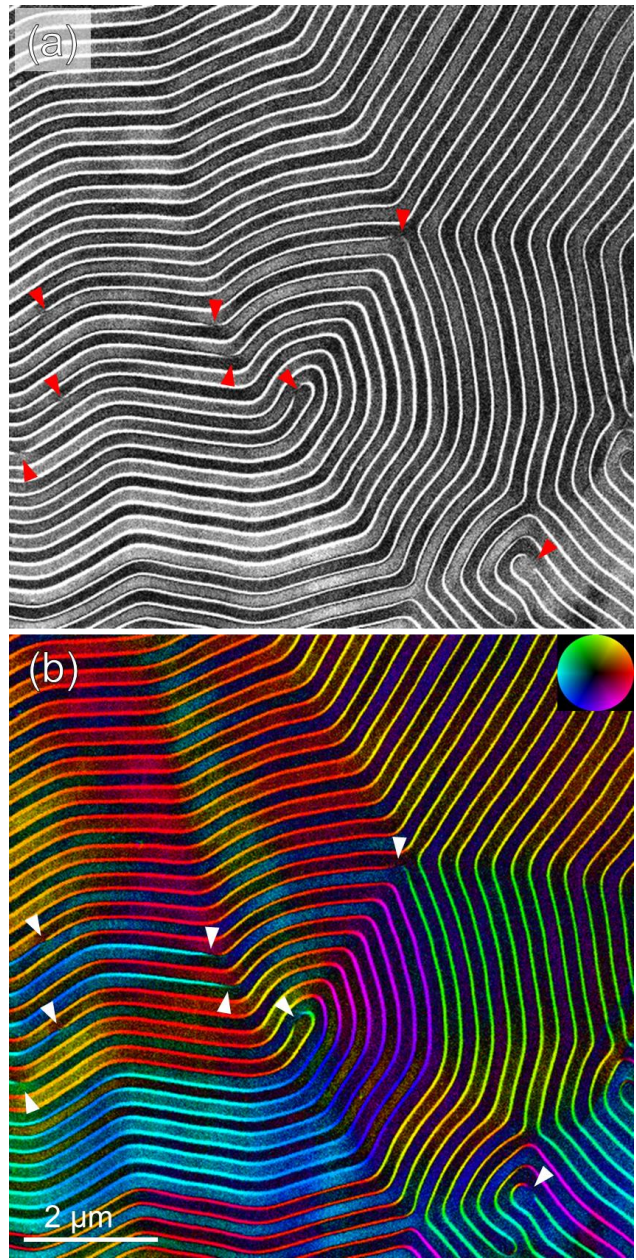

Fig. 2. (a) Field strength image of Bloch wall distributions in the $c$ plane of BFSMO. (b) Color map of the DPC STEM image of the same area as in (a). The arrowheads show the positions of the Bloch lines. The color wheels represent the magnetization direction.

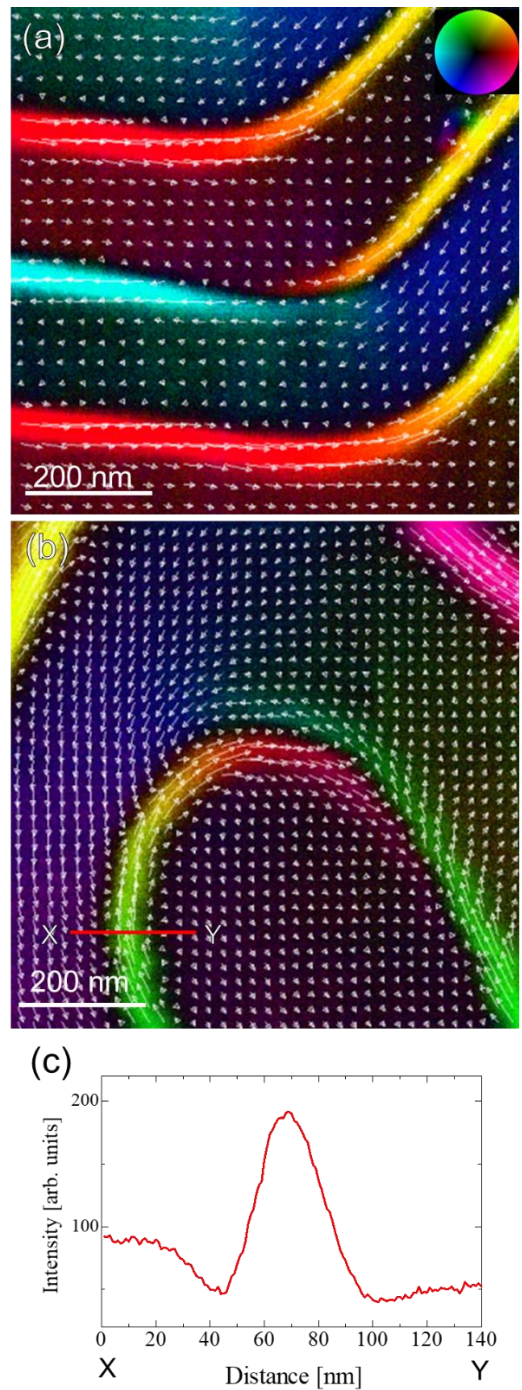

Fig. 3. High-magnification DPC STEM images (color maps) of (a) a tail-to-tail Bloch line and (b) a head-to-head Bloch line. The arrows show the magnitude and direction of magnetization. The color wheel represents the magnetization direction. Although the color is allocated in domains (due to a slight misalignment of *c* axis or the stray field from the Bloch walls), the magnitudes of magnetization (the sizes of arrows) in the magnetic domains are much smaller than those in the Bloch walls. (c) Intensity profile along X–Y in (b). The peak indicates that the wall width is approximately 50 nm.

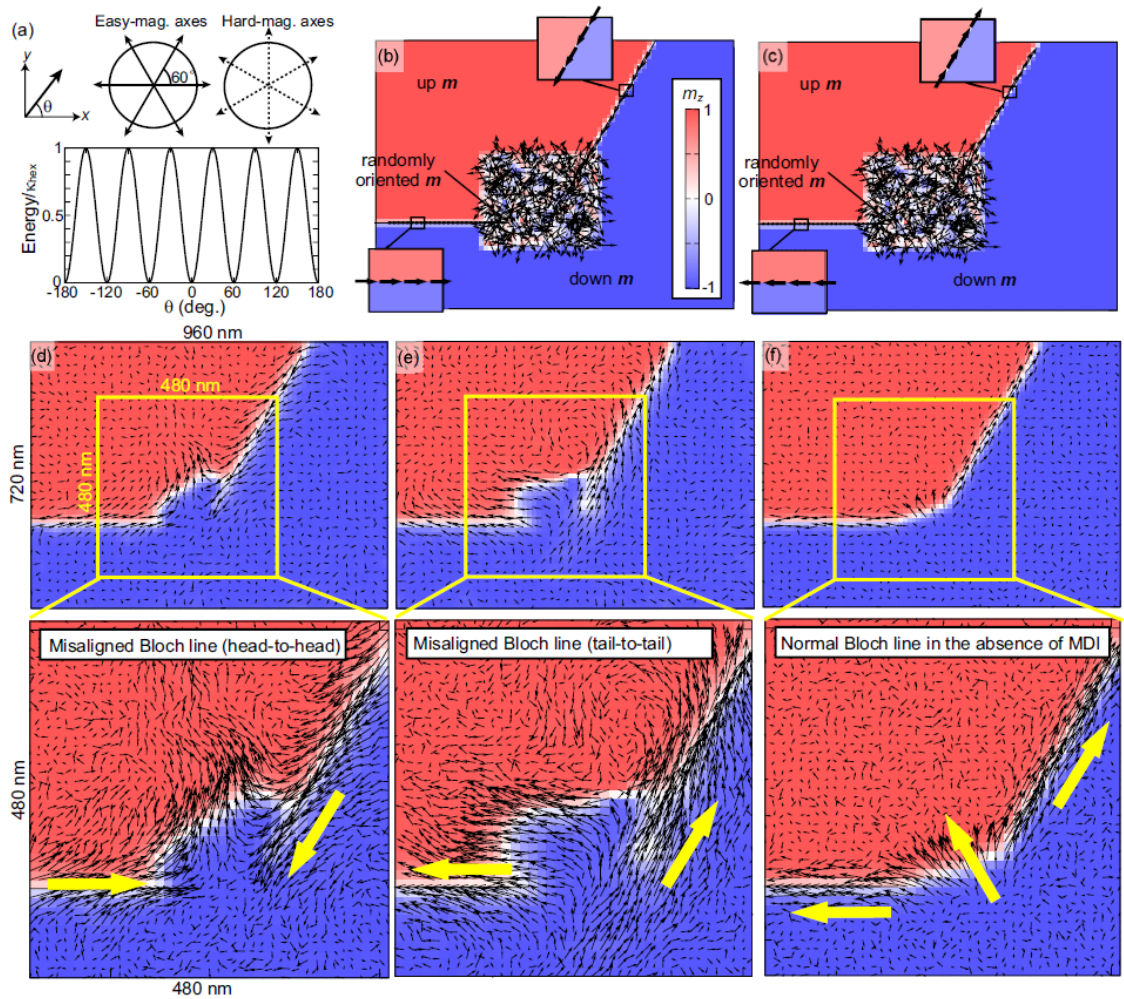

Fig. 4. (a) The in-plane six-fold magnetic anisotropy due to the hexagonal crystalline symmetry. The initial magnetization configurations for the replica-exchange Monte-Carlo technique in (b) head-to-head and (c) tail-to-tail configurations. (d), (e) Relaxed magnetic domain structures in each situation. (f) Relaxed magnetic domain structures without the dipole-dipole interaction. The Bloch walls are misaligned because of the dipole-dipole interaction, while the Bloch wall run along a line if the dipole-dipole interaction does not exist.

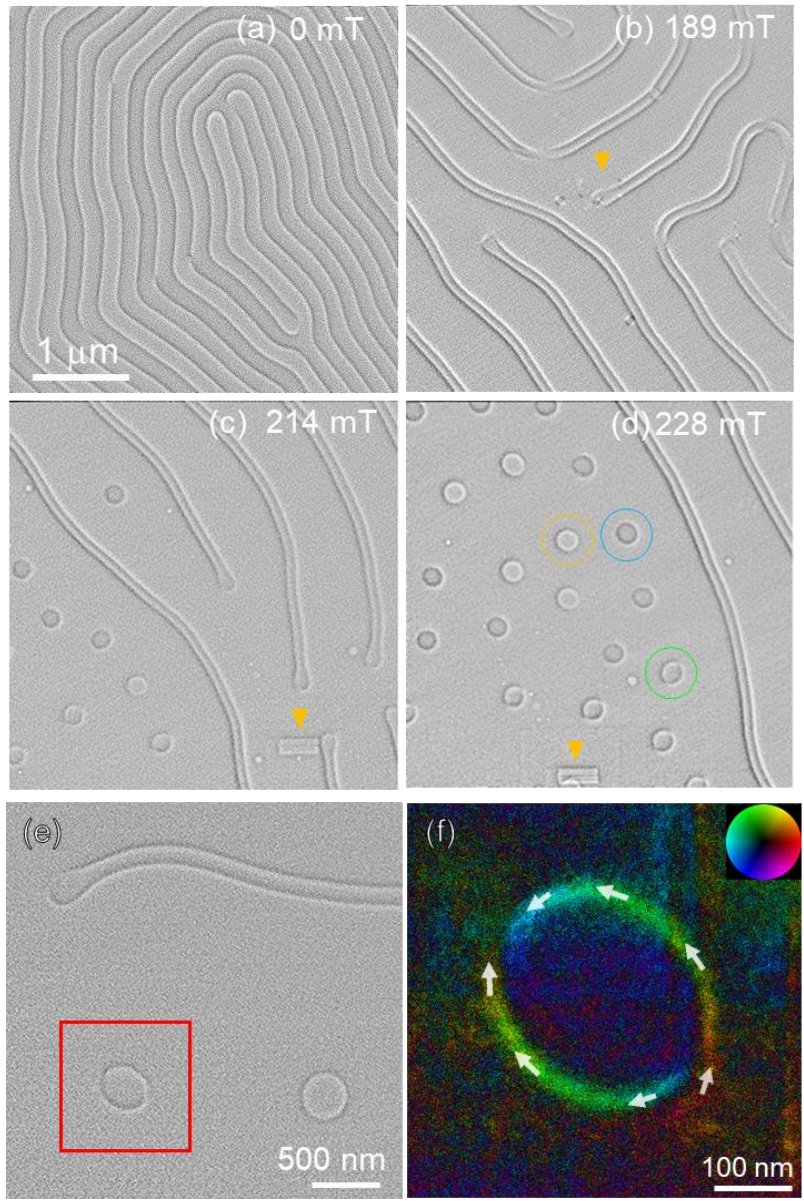

Fig. 5. (a)–(d) Changes in magnetic domains with increasing magnetic field. These images are depicted as rotation maps to show the helicity of magnetic bubbles. The yellow arrowheads indicate the same position in different images. The yellow, blue, and green circles show counterclockwise type-I, clockwise type-I, and type-II magnetic bubbles. (e) Magnetic domain-wall distribution (rotation map) under a magnetic field of ~228 mT in a narrow region. (f) High-magnification DPC STEM image (color map) of the type-II magnetic bubble marked by the red rectangular in (e). The white arrow shows the direction of magnetization in the Bloch walls.